# Physics Playground: Insights from a Qualitative-Quantitative Study about VR-Based Learning

Elena Battipede[1][0009-0006-3291-2663], Antonella Giangualano[2][0009-0001-7105-9988], Paolo Boffi*[2,3][0000-0003-4579-2606], Monica Clerici[2], Alessandro Calvi[3], Luca Cassenti[3], Roberto Cialini[3], Tristan Lieven Annemie Van Den Weghe[3], Loredana Addimando[1][0000-0002-4197-232X], Pier Luca Lanzi[3][0000-0002-1933-7717], and Alberto Gallace[2][0000-0002-4561-8926]

[1] Dipartimento formazione e apprendimento, Scuola universitaria professionale della Svizzera italiana, Piazza San Francesco 19, 6600 Locarno, Switzerland
[2] Università degli Studi di Milano-Bicocca, Piazza dell'Ateneo Nuovo 1, 20126 Milan, Italy
[3] Politecnico di Milano, Piazza Leonardo da Vinci 32, 20133 Milan, Italy
*corresponding author email: paolo.boffi@polimi.it

**Abstract.** *Physics Playground* is an immersive Virtual Reality (VR) application designed for educational purposes, featuring a virtual laboratory where users interact with various physics phenomena through guided experiments. This study aims to evaluate the application's design and educational content to facilitate its integration into classroom settings. A quantitative data collection investigated learning outcomes, related confidence, user experience, and perceived cognitive load, through a 2x2 within-between subjects setup, with participants divided into two conditions (VR vs. slideshow) and knowledge levels assessed twice (pre- and post-tests). A qualitative approach included interviews and a focus group to explore education experts' opinions on the overall experience and didactic content. Results showed an improvement in physics knowledge and confidence after the learning experience compared to baseline, regardless of the condition. Despite comparable perceived cognitive load, slideshow learning was slightly more effective in enhancing physics knowledge. However, both qualitative and quantitative results highlighted the immersive advantage of VR in enhancing user satisfaction. This approach pointed out limitations and advantages of VR-based learning, but more research is needed to understand how it can be implemented into broader teaching strategies.

## 1 Introduction

Virtual Reality (VR) represents a graphically-produced and three-dimensional digital model of an artificial environment, compiled in a computer to generate in the user a feeling of being immersed in a given place, thanks to the use of multisensory stimulation [11]. Immersive VR (IVR) systems, such as the Head-Mounted Display (HMD), perceptually surround the users and allow them to observe the Virtual Environment (VE) from a first-person perspective, increasing their sense of presence and the perception of the VE as real [14,40,48]. The growing interest towards VR in the educational field stems from the possibility for users to be exposed to stimuli that would be impossible, logistically complex or dangerous to reproduce in reality [40,24]. Many studies have been published about the efficacy of learning different disciplines in VR, such as History, Biology, Physics, Chemistry, or Science in general [15,24]. Among all disciplines, Physics is a suitable subject for VR, thanks to the possibility of creating virtual scenarios with altered physics constants or laws that cannot be changed on our planet, or the opportunity to visualize and interact with abstract concepts [24].

*Physics Playground* is an application for learning Physics in IVR, designed by a group of engineering students from Politecnico of Milan, within a collaborative project with Mind and Behavior Technological Center (MiBTec) of University of Milano Bicocca. The goal of our pilot study was to get an initial grasp at how *Physics Playground* can be improved, both in the design and didactical content, to potentially make it suitable for implementation in classrooms and educational settings in the future. Our main research questions were the following:

***Q.1*** *Is the Physics Playground application appropriate for teaching and learning, from a design and an educational content point of view?*

***Q.2*** *Is there a difference in learning physics in VR compared to learning the same concepts in a non-immersive and non-interactive modality such as a slideshow?*

We aimed to answer Q.1 through a qualitative study, by collecting qualified testifiers' opinions on the application, through interviews and a focus group at the Department of Teaching and Learning of SUPSI in Locarno, Switzerland. To answer Q.2, a second study took place at MiBTec laboratories, following a quantitative approach to collect data about the participants' overall experience and learning performances.

## 2     State of the Art - Learning in VR

After getting a broader perspective on the collected data, using a qualitative approach [10], we developed the following list of hypotheses to answer Q.2 quantitatively:

***H1****: Users who learned in VR will report higher user experience ratings than those who learned through a non-immersive and non-interactive modality such as a slideshow.*

Previous studies have observed high engagement and motivation in students when learning in VR and a higher rating of user experience as opposed to a traditional teaching modality, such as a slideshow [7,36].

***H2.1****: There will be no significant difference in post-learning performance between VR and slideshow learning.*

Humans learn not only by passively receiving information but also by actively engaging with the environment and receiving feedback from their actions – a principle supported by embodied cognition and constructivist theories. According to the former, knowledge is built on the interaction of body, mind, and environment, and it is grounded in experiences and the repetition of sensorimotor routines [47,49]. Thanks to its interactivity, VR technology allows users to manipulate the environment, acting directly on its components [19], and to perceive stimuli from different sensory modalities, which can also contribute to more memorable experiences [12]. Involving different senses in learning is particularly important in the constructivist approach, according to which learning is a continuous, active process where people build their representations of reality by extracting meaning from the sensorial experiences they have in the world, and by gradually updating their previous mental models based on feedback they get from the environment [14,40]. However, there are mixed findings on whether learning in VR is more effective than learning in a traditional modality: while some studies support this idea [1,29], others found better post-learning performances when viewing a slideshow compared to learning the same topics in VR [36,37]. In line with Jensen and Konradsen [22], we hypothesize that, despite their potential, using HMDs does not automatically cause learning to occur, as supported by other studies which found no differences between immersive and non-immersive learning modalities [7,23,9,31].

*H2.2: Those in the VR condition will feel more confident about their answers in the physics test after learning, compared to those who learned with more traditional methods.*

Measures of subjective confidence in one's own answers are often used in semantic memory tests [42,43]. Because VEs may be perceived as unfamiliar and uncomfortable, users may feel lower confidence when learning in VR compared to less immersive modalities [44]; on the other hand, users may feel more confident in what they have learned when they had a positive experience in VR [6]. Since we expect more positive evaluations of user experience for those in the VR condition, we also expect them to be more confident in their answers.

*H3: Perceived workload will be higher in VR than in slideshow learning.*

Although VR systems increase the auditory inputs and the visual field of view compared to 2D monitors, leading users to feel more present in the VE, this could also constitute a distraction and lead to a higher cognitive load in the VR condition. This could be due to the need to find relevant information among stimuli that are unrelated to the learning content [32,33], or the need to adapt to a new environment [35,39].

## 3 Materials and Methods

### 3.1 The Physics Playground Application

To address our research questions and verify the subsequent hypotheses, we designed an educational, fully immersive VR experience called *Physics Playground*, consisting of a virtual laboratory where users can engage in several physics experiments. The user can explore the VE by following an animated robot named Emanon, who serves as both a guide and a teacher throughout the experience. The environment simulates a Space Center, consisting of a set of office-looking rooms connected by a hallway. The four rooms are dedicated to fundamental physics concepts: gravity and friction, projectile motion, planetary orbits, and escape velocity. Emanon introduces the context and some theoretical aspects of each physical phenomenon through vocal dialogues paired with on-screen subtitles. Then, the user can actively modify the physics variables by interacting with a control panel, and conduct experiments to observe outcomes based on the updated values. In each room, the physics formulas behind the practical experiment are presented through blackboards to give interested users a deeper insight into the phenomena. To create a realistic setting and increase the provided immersion, we introduced a simple narrative: the user is asked to learn increasingly complex gravity-related concepts to be able to participate in a mission at a Mars space station. Throughout the experience, the user is given simple tasks to complete; however, each user receives standardized feedback independent of their task performance. Increasing the difficulty of the delivered content should provide a sense of challenge to the user, improving his or her engagement and motivation throughout the learning experience.

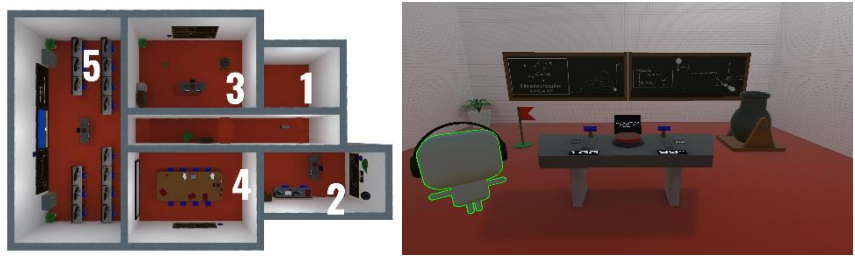

**Fig. 1.** On the left, the map of the environment, with the path the user should follow. On the right, one of the rooms, with all the common elements: Emanon, ready to speak, the control panel, and the blackboards.

**Learning Journey.** Throughout the entire experience, the user is required to follow Emanon's lead, as guided learning has been shown to enhance factual knowledge acquisition with respect to free-roaming structures [17]. The learning journey comprises a tutorial and four learning steps, one for each room. Each of them is designed to resemble an office to enhance the user's immersion and provide additional realism to the storyline (Fig. 1 shows the user's path). The experience begins with a tutorial where Emanon introduces the joystick controls for navigation and interaction. In the *Gravity room* (2), the user can experiment with basic gravity and friction concepts by increasing air pressure in a vacuum chamber and observing how two differently-shaped objects fall to the ground, either together or one after the other, depending on the amount of friction. In the *Cannon room* (3) participants can experiment with parabolic motion. The user is asked to hit the top of a red flag with the cannon's ball by varying both the cannon's angle and the launch speed. The *Planet room* (4) features three planets orbiting around the Sun. The user can modify the mass of each planet and the Sun, observing changes in mutual distances and rotational speed due to the change in gravitational attraction forces. The goal is to reach a balance between the planets' and the Sun's masses to maintain stable orbits. The final room, the *Control room* (5), simulates the control area of the Space Center. The user can manipulate the rocket's mass and the launch force to obtain an optimal balance to reach the escape velocity and make the rocket fly to Mars. The goal is to find an optimal combination of values to launch the rocket successfully.

**Environmental Interactions.** *Physics Playground* provides both experiment-specific and free-roaming-based interactions. During the experiments, the user can interact with Emanon and control panels. Emanon delivers educational content via audio, initiated by pointing at it with a ray and pressing a controller's button. After the explanation, the user can interact with a control panel to experiment with the introduced physical phenomenon. The control panel presents one or two levers (depending on the specific experiment), each corresponding to a physics variable, offering both ray-based and direct interactions, and a button that can be pushed to execute the experiment, with only direct interaction allowed. In addition to the experiment-related interactions, we decided to make most of the objects of the environment interactable: the user can grab almost anything (books, ring binders, ...) in the rooms, with both ray-based and direct interactions, and either move them or throw them around. Though not strictly related to the educational aim of the experience, these interactions enhance the realism and immersion of the environment.

### 3.2   Methods

**Qualitative Study.** Involving teachers and experts in education in the design of educational tools and activities can be beneficial, as their attitudes and points of view can help in finding the most effective ways to structure the learning environments [27]. So, we conducted a focus group with 7 students who are training to become high school teachers and interviews with 6 Professors and experts in using Mathematics, Science, Technologies, and Media in education, recruited from the Media and STEM Laboratory at the Department of Teaching and Learning of SUPSI. After trying the application on a Meta Quest 2 HMD connected to a gaming laptop, participants told their first general impressions of the VR application. Questions for the interviews and the focus group were based on Loorbach et al.'s [30] Reduced Instructional Materials Motivation Survey (RIMMS). This phase of the study aimed to collect the participants' opinions, get more details about their suggestions, or touch topics such as the likeability of the experience, the appropriate target age of users, and potential changes in the graphical design to improve immersion.

**Quantitative Study.** Participants were randomly assigned to either the experimental condition (N = 22), which consisted of learning through our VR application, or a control condition (N = 22), which consisted of learning the same physics topics through a slideshow. Before the learning phase, all participants filled out a pre-test consisting of 19 multiple-choice physics questions, 7 of which unrelated to the main topics of the application to prevent participants from focusing only on the topics presented in the learning phase. Each question was paired with an item asking the confidence level in answering the question, from 0 (Not sure at all) to 100 (Completely sure). Then, those in the experimental condition started the application. Once completed, they had the possibility to explore the VE freely, until they reached a total immersion time of 20 minutes, kept constant between participants. Conversely, those in the control group followed a series of slides presented in both visual and audio format using an automated voice-over-text tool. The slideshow took about 5 minutes, but each participant was allowed to return to previous slides or read them as long as needed. The written texts in the slides followed the script of the application's robot without the storytelling and emotional aspects.

The participants were then asked to complete the second part of the questionnaire, including a physics test with the same questions as the pre-test in randomized order, the level of confidence related to each question, the short version of the User Experience Questionnaire (UEQ; [28,45,46]) and the Task Load Index (TLX; [20]) to measure perceived workload. Additionally, other measures were taken with an exploratory aim: System Usability Scale (SUS; [8]), Virtual Reality Sickness Questionnaire (VRSQ; [25]), Presence Questionnaire [50], and one embodiment-related item adapted from the Inclusion of Other in Self (IOS) scale [3] were measured to ensure that the application was working as expected.

## 4      Results

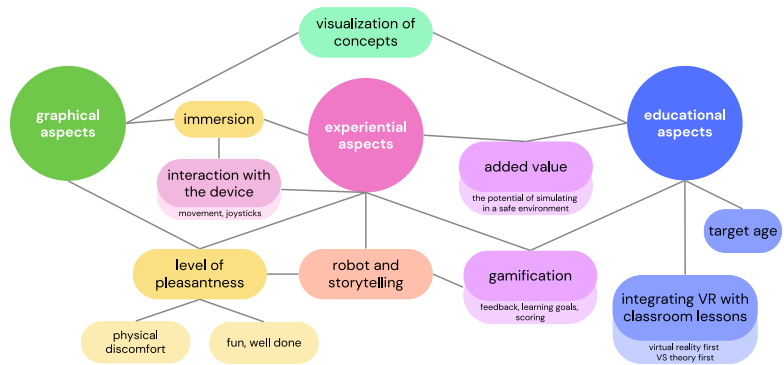

**Fig. 2.** Schematic representation of themes from interviews and focus group

For the qualitative study, themes emerged from the thematic analysis (Fig. 2) will be included in the discussion paragraph. In the quantitative study, the sample consisted of 44 participants (22 males, 21 females, 1 non-binary, aged between 19 and 34).
**H1.** A preliminary frequency analysis of the UEQ answers, including those of both groups, revealed that the Pragmatic Quality (PQ), Hedonic Quality (HQ), and Overall scores were above a neutral evaluation, regardless of the experimental condition (Fig. 3). An independent samples Welch's t-test revealed significant differences between conditions in HQ ($t(30.8) = -6.957$, $p < .001$) and Overall score ($t(31.2) = -4.481$, $p < .001$), with the VR group exhibiting higher evaluations of both HQ ($M = 2.22$, $SD = 0.67$) and Overall score ($M = 2.04$, $SD = 0.54$) compared to the control group (HQ: $M = -0.0114$, $SD = 1.34$; Overall score: $M = 0.9041$, $SD = 1.06$). No significant differences were found for PQ. We also found significant positive correlations between HQ and the sense of presence score ($r = 0.489$, $p = 0.021$), Overall score and the sense of presence score ($r = 0.541$, $p = 0.009$), and between PQ and SUS scores ($r = 0.483$, $p = 0.001$).

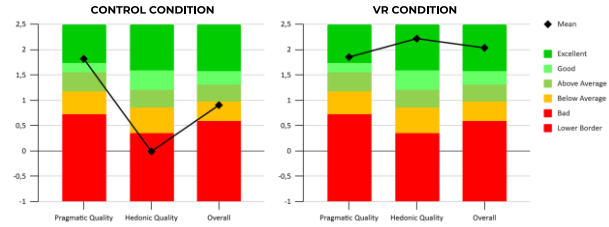

**Fig. 3.** UEQ Benchmark comparison using Schrepp et al.'s [45,46] tool.

**H2.1.** Scores for the pre-test and post-test (Fig. 4a) were determined by assigning 1 point to each correct answer, excluding questions unrelated to the application's contents. An independent samples Student's t-test found no differences in the pre-test scores between conditions ($t(42) = -1.38$, $p = 0.176$). A Wilcoxon signed-rank test showed a significant negative difference between pre-test and post-test scores ($W = 8.00$, $p < .001$). This result was consistent across both conditions: control ($W = 0.00$, $p < .001$), and VR ($W = 5.00$, $p < .001$). A Mann-Whitney U test revealed a significant difference in post-test scores ($U = 152$, $p = .029$), with the control group showing a higher mean ($M = 10.30$, $SD = 1.42$) compared to the VR group ($M = 9.23$, $SD = 2.02$). To represent the difference between each participant's post-test and pre-test scores, we computed a new variable called *delta_test*, with positive values corresponding to a better performance in the post-test. A Mann-Whitney U test showed that the means of this variable were significantly different between conditions ($U = 130$, $p = .008$), with the control condition having a higher mean ($M = 4.27$, $SD = 2.35$) than the VR condition ($M = 2.32$, $SD = 2.03$). A mixed ANOVA 2x2 found a significant main effect of learning on test performance ($F(1, 42) = 98.80$, $p < .001$) but no significant difference between conditions ($F(1, 42) = 0.0188$, $p = 0.892$). We found a significant interaction effect between condition and test performance ($F(1, 42) = 8.69$, $p = 0.005$).

**H2.2.** We found no differences between conditions in confidence levels for pre-test and post-test questions, even when considering only the questions related to the application's topics. Paired samples Student's t-tests showed significant differences in confidence between pre-test and post-test with: all questions ($t(43) = -14.39$, $p < .001$), topic-related questions ($t(43) = -13.30$, $p < .001$), and unrelated questions ($t(43) = -4.56$, $p < .001$).

**H3.** The independent samples Student's t-test did not reveal a significant difference in TLX scores (Fig. 4b) between conditions ($t(42.0) = 1.94$, $p = 0.060$). Among the measures taken at an exploratory level, it is worth mentioning the SUS-related results. A Mann-Whitney U test revealed a significant difference between conditions ($U = 135$, $p = .019$), with the control group having a higher mean ($M = 84.6$, $SD = 10.2$), than the VR one ($M = 78.6$, $SD = 9.28$). Finally, SUS scores positively correlate with Presence scores ($r = .590$, $p = .004$).

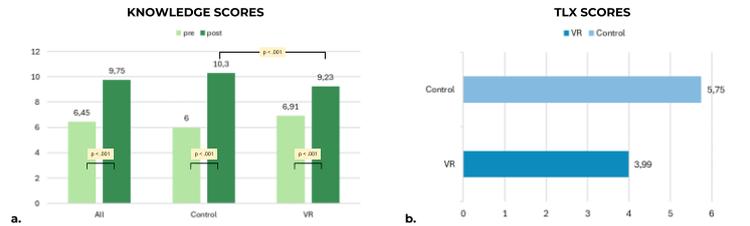

**Fig. 4.** (**a**) Pre- and post-knowledge scores (**b**) TLX scores

## 5   Discussion

**H1: confirmed.** The VR system outperformed traditional learning in hedonic quality and overall scores, highlighting its potential to enhance visual appeal, innovation, and positive emotions. Additionally, VR was perceived by participants to be just as valid as a more familiar tool like slideshows in all those aspects that focus on practical use, such as ease of use, efficiency, effectiveness in achieving a goal, and clarity. Moreover, higher ratings of the hedonic aspects and overall user experience were associated with higher experienced sense of presence, reinforcing the immersive advantage of VR in enhancing user satisfaction. Some participants from the qualitative study also described the VR immersion as a positive experience, while others found the environment cold and mazy. In order to improve the experience, they also suggested improving the graphical aspects of the VE by inserting more immersive elements such as a cosmic space with planets and the Sun in the center.

**H2.1: not confirmed.** Results showed similar baseline knowledge and a general knowledge increase, regardless of the learning condition. However, contrary to the hypothesis, the control group outperformed the VR group, suggesting that traditional slideshow learning was slightly more effective in enhancing physics knowledge. VR seems yet unable to surpass traditional learning methods in improving factual knowledge. Introducing VR into education brings benefits in terms of engagement and motivation, but it should be integrated into broader strategies rather than used as a standalone method [41,38,2]. This emerged from the qualitative study as well, in which participants agreed that the VR application is currently not usable on its own and needs to be theoretically introduced or explained by teachers. Moreover, they suggested leveraging graphical design to visually represent abstract theoretical concepts (e.g., by making the boards interactive).

**H2.2: not confirmed.** Both groups gained confidence after the learning experience, showing that traditional and VR-based learning environments can effectively increase students' confidence. The learning condition did not affect confidence levels: despite the novelty of the VR system for most participants [26,35], their confidence was not negatively impacted and VR proved to be just as effective in boosting confidence as traditional learning methods, which students are more familiar with.

**H3: not confirmed.** Contrary to our expectations, there were no differences in perceived cognitive load between the VR and control groups. These results seem to suggest that the extraneous cognitive load generated by VR systems, due to exposure to a sensory-rich environment and the need to learn new controls, could be attenuated by the appealing features of the system itself. According to the Cognitive-Affective Theory of Learning with Media [32] the sensory richness and interactivity of immersive technologies enhance intrinsic motivation while reducing the mental effort required for learning. Additionally, recent studies [16,32] explored the connections between cognitive load and learners' motivation to better understand their influence on learning performance, finding that engagement and intrinsic motivation can reduce perceived cognitive load. Despite the unfamiliar and highly stimulating multisensory environment, the motivation and engagement fostered by VR can make learning feel less mentally demanding [32].

Our data collection revealed interesting usability results. Despite having scored above the usability benchmark [4,5,18] with high scores, demonstrating good usability, the VR application was rated with significantly lower SUS scores than the slideshow. We attribute this result to participants' higher familiarity with slideshows, a tool commonly used in learning contexts. However, increasing the frequency of VR use might lead to higher learning results and lower perceived cognitive load.

## 6    Conclusion

As emerged from our qualitative-quantitative study, *Physics Playground* provides an overall positive experience, but more needs to be investigated to understand how to improve its educational impact and ability to transmit knowledge.

### 6.1    Limitations and Future Research

Participants of the qualitative study suggested an age-range between middle school and first years of high school as a target for Physics Playground. Therefore, the next step will be integrating Physics Playground within the traditional teaching approach to maximize its educational and learning potential in a more ecological setting, testing the suggested target.

An additional issue regards the lack of follow-up assessments evaluating long-term knowledge adequately in VR [33]. Future studies should implement additional assessments to evaluate physics content retention over a longer time span and investigate the possible influence of sense of presence and embodiment.

Another promising approach is personalizing the application's contents to align with students' learning styles [13], which can influence the subjective sense of presence and cognitive load in the learning process [21]. Therefore, future versions of Physics Playground should include new features such as a personalized feedback system, guided assistance, and gamification techniques. This also emerged from the focus group and interviews with educational experts, who mentioned the incorporation of gamification elements (progress bars, a scoring system, prizes, a list of the user's previous attempts and more personalized feedback), which are effective in boosting engagement and motivation towards the accomplishment of learning goals [34].

Finally, it is important to focus on the added value that Physics Playground can give to the learning experience compared to traditional methods: according to the qualitative study participants, the VR experience is helpful in simulating experiments that cannot be done in class due to practical or safety reasons [24]. Many subjects declared seeing potential in the application, while others were dubious about it for practical and economic reasons. Cost remains a significant barrier to VR adoption in schools due to software updates, content creation, and instructor training costs [22]. We are already considering this aspect, implementing and testing a less-costly Desktop VR version of the application. The final goal of this project, as emerged from our qualitative results, seems to be finding the "sweet spot" between engaging and effective teaching.

**Acknowledgments.** We are thankful to Prof. Giancarlo Gola from SUPSI, his students, and all the participants who voluntarily took part in the data collection.

**Disclosure of Interests.** The authors have no competing interests to declare that are relevant to the content of this article.